# Free Evolution of Superposition States in a Single Cooper Pair Box

[1]A. Guillaume, [2]J. F. Schneiderman, [1,3]P. Delsing, [2]H.M. Bozler and [1]P. M. Echternach

[1]Jet Propulsion Laboratory, California Institute of Technology, Pasadena, CA 91109

[2]University of Southern California, Department of Physics and Astronomy, Los Angeles, CA 90089-0484,

[3]Chalmers University of Technology, Microtechnology and Nanoscience, MC2, 412 96 Göteborg, Sweden

We have fabricated a single Cooper-pair box (SCB) in close proximity to a single electron transistor (SET) operated in the radio-frequency mode (RF-SET) with an inductor and capacitor lithographed directly on chip. The RF-SET was used to measure the charge state of the SCB revealing a 2$e$ periodic charge quantization. We performed spectroscopy measurements to extract the charging energy ($E_C$) and the Josephson coupling energy ($E_J$). Control of the temporal evolution of the quantum charge state was achieved by applying fast DC pulses to the SCB gate. The dephasing and relaxation times were extracted from these measurements.

03.67.Lx, 74.50.+r, 85.25.Cp, 85.35.Gv

In recent years, tremendous progress has been achieved in various superconducting circuit designs for quantum computing[1,2,3,4,5,6,7]. The main difference between these implementations relies on the relative importance of the charging energy and the Josephson coupling energy, defining three varieties of qubit: flux, phase, and charge. When the charging energy dominates over the Josephson coupling, the charge on the island is used as the quantum degree of freedom. Since the seminal work of Nakamura et al.[1] on the first control of coherent charge states, other experiments estimate the relaxation times[2] and observe the temporal evolution of the charge state in a single Cooper-pair box (SCB)[3,7]. In the other limits, control in the time domain has been achieved for the flux state[4] and for the phase qubit[5,6]. While the NEC group[1] probed the charge state of the SCB with quasiparticle tunneling through a highly resistive junction, an alternative approach has been explored which employs a radio-frequency single electron transistor[8] (RF-SET) for faster readout[9]. The speed of the RF-SET together with its sensitivity should allow for single-shot measurement of the charge state. The temporal control of the SCB charge state is observed with an RF-SET and the relaxation and dephasing times measured[3]. Other efforts focus on increased integration by designing the ensemble readout-qubit in a more embedded system. The Saclay group enclosed an SCB qubit and a large Josephson junction in a loop for integrated readout[7], while the Delft[4] group has joined their qubit with a DC squid readout system.

The current work presents the first results obtained with a circuit in which all the components of the faster RF-SET readout are fabricated on the same chip as the SCB qubit. We report on the time control of the charge states of the SCB achieved by means of fast DC pulses applied to the SCB gate. Our samples, as shown in Figure 1a, consist of an SCB and an RF-SET fabricated on a quartz substrate using electron beam lithography and standard shadow mask techniques[10]. The LC tank circuit and other contact structures were fabricated using an Al/Ti/Au trilayer structure[11] allowing for tailoring of the transition temperature (~0.5 K for the present samples). By selecting a transition temperature well below that of pure Aluminum (~1.2 K), the LC circuit and contacts act as traps for quasiparticles[12] that otherwise degrade the coherence time of the qubit.

Our SCB consisted of two small Al/AlO$_x$/Al tunnel junctions connecting a reservoir to a small island. The two junctions in parallel formed a SQUID loop allowing for control of the Josephson coupling energy, $E_J = E_J^{max} |\cos(\pi \Phi / \Phi_0)|$, by threading the loop with a magnetic flux (where $\Phi_0 = e/2\hbar$ is the flux quantum) generated by a pair of Helmholtz coils placed outside the cryostat. The SET island was fabricated in close proximity to that of the SCB, allowing for fast measurement of the charge state through capacitive coupling while operating the SET in the RF mode. In addition, two gate leads were placed near the island; the slow SCB gate was used to sweep the charge on the island with a low frequency ramp and the fast SCB gate was used to apply fast DC pulses to manipulate the quantum states of the SCB. The sample was thermally connected to the mixing chamber of a dilution refrigerator and all the results reported here were obtained at a base temperature of about 15 mK.

The free evolution of a superposition state generated by controlling the charge on the island via the fast SCB gate is the basis for operating an SCB as a qubit.[13] The SET island was fabricated very close to the SCB island so that a charge appearing on the SCB island was coupled to the SET via a capacitance $C_C$ as shown in Figures 1a and b. Charges on the SCB were coupled to the SET island with a ratio $C_c/C_\Sigma$ and modulated the reflected power on the RF-SET. Since the total capacitance of the box island $C_\Sigma$ was small, the associated charging energy $E_C=e^2/2C_\Sigma$ dominated over the Josephson energy (and the energy of thermal fluctuations) so that the number of Cooper-pairs on the island was the degree of freedom that best described the quantum state of our system. The slow SCB gate controlled the charge induced on the island as $N_g=C_gV_g/2e-n_{off}$ (in units of $2e$), where $n_{off}$ represents an uncontrolled but fixed charge offset. The tunneling matrix element expressing the coupling between two different charge states has an amplitude $E_J/2$. At low temperatures the Josephson energy, $E_J$, is[14][15] $(R_Q/R_N)(\Delta/2)$ where $R_N$ denotes the normal state resistance of the junctions, $R_Q=h/(4e^2)$ the quantum resistance, and $\Delta$ the superconducting gap. Provided $k_BT<E_J<E_C<\Delta$, the SCB can be considered an artificial two-level system. In the Cooper pair charge basis $\{|0\rangle,|1\rangle\}$ its Hamiltonian can be written:

$$H = -\frac{\vec{\sigma}\cdot\vec{h}}{2}$$

(1)

describing a pseudo-spin $\vec{\sigma}$ in an effective magnetic field $\vec{h}=[E_J,0,4E_C(1-2N_g)]$. The vector $\vec{\sigma}$ components are the usual Pauli matrices. The energy-level diagram is represented on Figure 2 versus the normalized gate charge $N_g$. The Josephson coupling creates an energy-splitting $E_J$ for a gate charge equal to half a Cooper pair. The expectation value of the excess charge on the island is calculated with the eigenvectors of Hamiltonian (1) and is represented on Figure 2. This average charge as a function of $N_g$, known as the Coulomb Staircase[16], is the experimental quantity that we measured. In order for this staircase to be $2e$ periodic, the condition $\tilde{\Delta} < E_C - E_J/2$, where $\tilde{\Delta}$ is the even-odd free energy difference,[17] must apply. If it does not, a short step characteristic of quasiparticle poisoning will appear in between the $2e$ periodic steps. The effect of a non-adiabatic DC pulse on the qubit gate is indicated on Figure 2. The rising part of the pulse brings the system from the ground state to the degeneracy point, which means that it will induce a $\pi/2$ rotation of the effective magnetic field, $\vec{h}$. At this point, the effective magnetic field has only a component on the x-axis, $h_x = E_J$, and the quantum state precesses around the magnetic field with an angular frequency $E_J/\hbar$ throughout the duration of the pulse $\Delta t$. The fall of the pulse then performs another $\pi/2$ rotation of the effective magnetic field $\vec{h}$. The time evolution operator $U(\Delta t) = \exp(i\alpha\sigma_x)$, with $\alpha = E_J \Delta t / 2\hbar$, applied to the ground state describes this precession of the quantum state. The principle of controlling the weight of each quantum state in the superposition of states by adjusting the time $\Delta t$ is the basis for a single qubit operation.

We have studied two samples, hereafter referred to as Samples 1 and 2. To measure the Coulomb staircase, we swept the slow qubit gate at low frequency and monitored the induced charge change on the SET, which was continuously operated in the RF mode and biased at the double Josephson quasiparticle peak[18] to minimize the back-action of the SET on the SCB. Each trace represents an average over 12,800 sweeps. The resulting staircase for sample 1 measured with the slow SCB gate swept at $1.13 \times 10^4$ $e$/s is represented by the thick line on Figure 3 and displays the periodicity characteristic of the 2$e$ quantization of the charge on the SCB. However, the small step visible at the end of the long step indicates the presence of quasiparticles. It is worth noticing that this feature became more pronounced and the staircase nearly $e$ periodic, when we swept the gate at a slower rate of 170 $e$/s (dashed line in Figure 3).

We then performed spectroscopy of the two-level system by sending a continuous microwave signal to the fast SCB gate while sweeping $N_g$ with the slow SCB gate. When the microwave energy $h\nu$ was resonant with the energy difference between charge states, a peak and a dip appeared at a distance $\Delta N_g = \sqrt{(h\nu)^2 - E_J^2}/8E_C$ from the degeneracy point, $N_g$=0.5. The grey line on Figure 3 depicts the effect of 30 GHz irradiation on the Coulomb staircase. Our values and theoretical fits for $\Delta N_g$, corresponding to the distance between the peak and the dip, are shown in the inset of Figure 3. For sample 1 we extract the charging energy $E_C$=2.47 K and the Josephson energy $E_J$=0.17K ($\pm$ 0.44K) from a fit of the experimental data. The value of the Josephson energy is only indicative since the data could also be extrapolated linearly to zero and the confidence interval very large. We can compare this value with the estimate for the Josephson energy $E_J^{max}$ of 152$\pm$22 mK obtained with the Ambegaokar-Baratoff formula[14][15] using the values $\Delta$=2.3$\pm$0.1 K and $R_N$=195$\pm$5 k$\Omega$ extracted from the IV characterization of the SET and assuming the sizes of the junctions to be the same (within 10%) for the SET and the SCB as designed. This data was taken while operating with no magnetic field, although no special care was taken to shield the sample from stray fields like that of the earth. Spectroscopy measurements were also performed on sample 2 with an applied magnetic field of 5Oe, and the fit of the experimental results yields $E_J$=0.27K (±0.12K) and $E_C$=1.03K. For Sample 2, the fit to the Ambegaokar-Baratoff formula yields an $E_J^{max}$ of 571±82 mK from an $R_N$ of 104±3 kΩ. The value of $E_C$ is reasonable since by design the area of the junctions is twice as large as the areas of the junctions in sample 1. The value of $E_J$ is smaller than maximum predicted value due to the presence of the magnetic field.

The $E_C$ extracted from the fit was larger than the superconducting gap for Sample 1 and the presence of the small step in its Coulomb staircases indicated the presence of unbound quasiparticles on the SCB and thus the odd-even free energy, $\tilde{\Delta}$, was smaller than the difference $E_C - E_J/2$. Our attempt to control the quantum state proved unsuccessful due to this quasiparticle poisoning, and we decided to remedy the situation by designing Sample 2 with larger tunnel junctions decreasing $E_C$ and increasing $E_J$ as described above. Indeed, the staircases obtained for Sample 2 were always 2e periodic (for all gate ramp frequencies) and we were able to see the temporal evolution induced by pulses.

We performed two kinds of time-resolved experiments on Sample 2. First, we applied trains of DC pulses for different pulse widths $\Delta t$ with a fixed repetition time $T_r$ between two pulses on the fast gates of the SCB. The effect of a single pulse is schematically described in Figure 2. We swept the voltage on the slow SCB gate at a low frequency of 3816 $e$/s while simultaneously applying the train of DC pulses on the fast SCB gate with a repetition rate $T_R$=30 ns that maximized the amplitude of the charge oscillation while we varied $\Delta t$ between 100 ps and 2500 ps. We subtracted from each individual staircase the average of all staircases (for different pulse widths). The signal remaining is therefore approximately the deviation from the ground state staircases induced by the application of the pulses.

The inset of Figure 4 shows the charge oscillations observed in the middle of the steps versus the pulse width. The magnitude of the first oscillation was approximately 0.7±0.1 $e$ and the dominant period was 109 ps corresponding to a Josephson energy of 440 mK in fair agreement with the $E_J^{max}$ value of 571±64 mK deduced from the SET parameters and the range 150 to 390mK estimated from the spectroscopy fit. With a period of ~109 ps, our data correctly extrapolated to a minimum at zero pulse width, where the system should remain in the ground state. The excitation by the pulses can be considered non-adiabatic because the oscillation period of 109 ps is larger than the 60 ps rise-time of our pulse generator. The oscillations show beating behavior characteristic of two superimposed sine waves with different frequencies. One possible explanation for this behavior is that there were unwanted reflections on the fast SCB gate line that deformed the DC pulses in different ways depending on their width. Decay in the amplitude of these oscillations gave a lower-bound estimate for the relaxation time $T_2$ in our system of roughly 1 to 2ns. We compared this value of $T_2$ with the one obtained by studying the spectroscopy line shape[19,2]. Extrapolating the half width at half maximum (HWHM) of the spectroscopy peaks, expressed in frequency through the relation $\delta\omega=(1/\hbar)(dE_{01}/dn_g)\delta n_g$, to the limit of zero microwave power yields the decoherence rate $1/T_2$ of a two-level system. The power dependence experiment was taken at a magnetic field of 3.8 Oe (different than the one used for the data shown on Figure 4) and resulted in a $T_2$ of 120 ps. As was argued by Lehnert et al [2] this measurement is a worst case estimate, which is nicely confirmed by the longer time estimated form the temporal evolution.

Once we knew the oscillation time, we performed a second type of experiment where we studied the relaxation from a given quantum state to the ground state. We chose this state by fixing the pulse width at 250 ps and varying the repetition time from 20 ns to 400 ns (Figure 4) and observed a non-exponential decay of the oscillation amplitude. However, if we assume that the probability to be in the excited state decays exponentially over a time scale $T_1$ and average over one pulse cycle, we can use the expression derived by T. Duty et al.[3]: $\Delta Q(T_R) = 2n_0(T_1/T_R)(1-e^{-T_R/T_1})/(1+e^{-T_R/T_1})$. A fit of the data to this formula yielded a relaxation time $T_1 = 52 \pm 16$ ns. Estimates of the relaxation time caused by quantum fluctuations of a 50 Ω environment capacitively coupled to the qubit yield results[20] that are between two and three orders of magnitude higher than our experimental result. This mechanism is therefore probably not the main source of energy relaxation in our system. A previous measurement of $T_1$ with a different experimental technique[2], using spectroscopy, but in the same type of SCB and RF-SET system, lead to a much longer relaxation time of 1.3 μs. This raises the question whether this difference stems from a different measurement scheme or from a different electromagnetic environment.

The values of $T_1$ and $T_2$ found in our experiment are similar to the ones found in reference[3] with similar experimental techniques.

In conclusion, we have fabricated Single Cooper Pair Boxes in conjunction with Single Electron Transistor readouts. We have placed this artificial two-level system in a superposition state by rapidly changing the Hamiltonian by way of a voltage pulse. By varying the duration of the pulse we followed the temporal evolution of this superposition state and measured a decoherence time and a relaxation time. With further improvements on the measurement set up we expect to obtain better control of this temporal evolution of the superposition states and be closer to a physical realization of a quantum bit, one step on the way to realizing a functional quantum computer.

We would like to thank Rich E. Muller for the electron-beam lithography and Alexander Korotkov and Rusko Ruskov for useful discussions. We performed this work at the Jet Propulsion Laboratory, California Institute of Technology, under a contract with the National Aeronautics and Space Administration. This work was partially funded by the Advanced Research and Development Activity (ARDA), the National Security Agency (NSA) and the National Science Foundation under contract number 0121428.

**Figure 1: (a) Scanning electron micrograph (SEM) of Sample 1 (b) Schematic corresponding to SEM picture. The tank circuit surrounded by the dashed line is located on the same chip, but outside the field of view on Figure 1 (a).** $C_C$

capacitively couples the SCB to the SET. The loop geometry of the SCB allows for flux tuning of the Josephson energy, $E_J = E_J^{max}|\cos(\pi\Phi/\Phi_0)|$.

Figure 2: Energy diagram in the two-level approximation. The Josephson coupling energy lifts the degeneracy at $N_g$=0.5 and induces a splitting, $E_J$. The rising part of the square pulse brings the system from the ground state to the degeneracy point (plain arrow) where the system oscillates between the two states $(|0\rangle \pm |1\rangle)/\sqrt{2}$ for the time $\Delta t$ set by the pulse width. When the pulse is removed and the charge is brought back to the starting point (dashed arrow), the quantum state evolves freely. The expectation value of the charge on the SCB as a function of gate voltage has the characteristic shape of a Coulomb staircase (dotted line).

**Figure 3**: The Coulomb staircase taken with Sample 1 without excitation, with a sweep rate of $1.13 \times 10^4$ $e$/s (dark line) and with a rate of 170 $e$/s (dashed line). The gray line shows the effect of continuous irradiation of the SCB at 30 GHz. Inset: Excitation frequencies corresponding to the differences $2\Delta N_g$[2e] for Sample 1 (closed symbols) and for Sample 2 (open symbols). On the vertical axis, the lower and upper triangles represent estimates for $E_J$ based on sample parameters for Samples 1 and 2, while the square represents the coherent oscillation frequency and the diamond represents the best numerical fit in Sample 2 corresponding to $E_J$ =273 mK ±120 mK.

**Figure 4**: Results of pulse experiments on Sample 2. (a) For a fixed repetition time of 30 ns, we varied the pulse width and observed charge oscillation at the center of the Coulomb staircase. The oscillation has a period of ~109 ps and is damped in ~2 ns ($T_2$). (b) For a fixed pulse width of 250 ps, we measured the charge oscillation amplitude decrease with the repetition time. From the fit (dotted line) we extracted an energy relaxation time of $T_1$=52±16 ns.

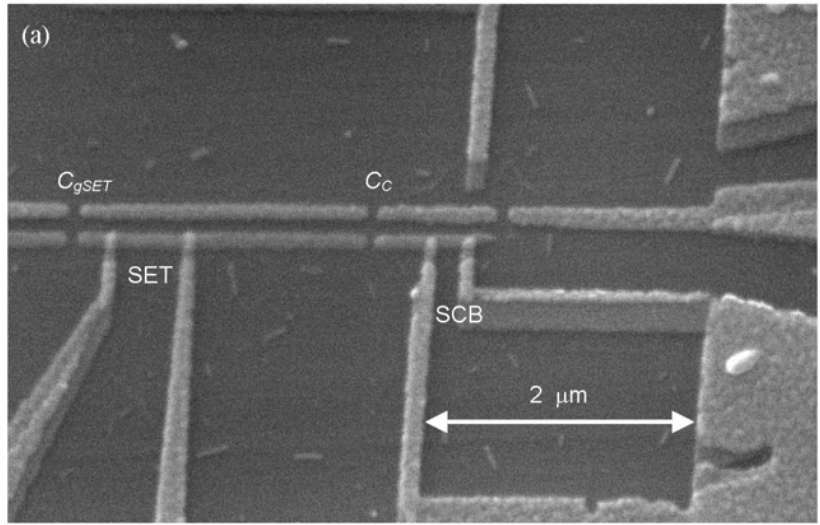

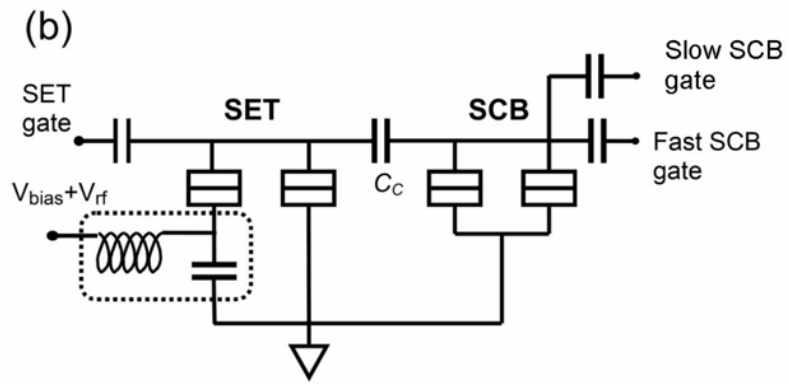

**Figure 1.**

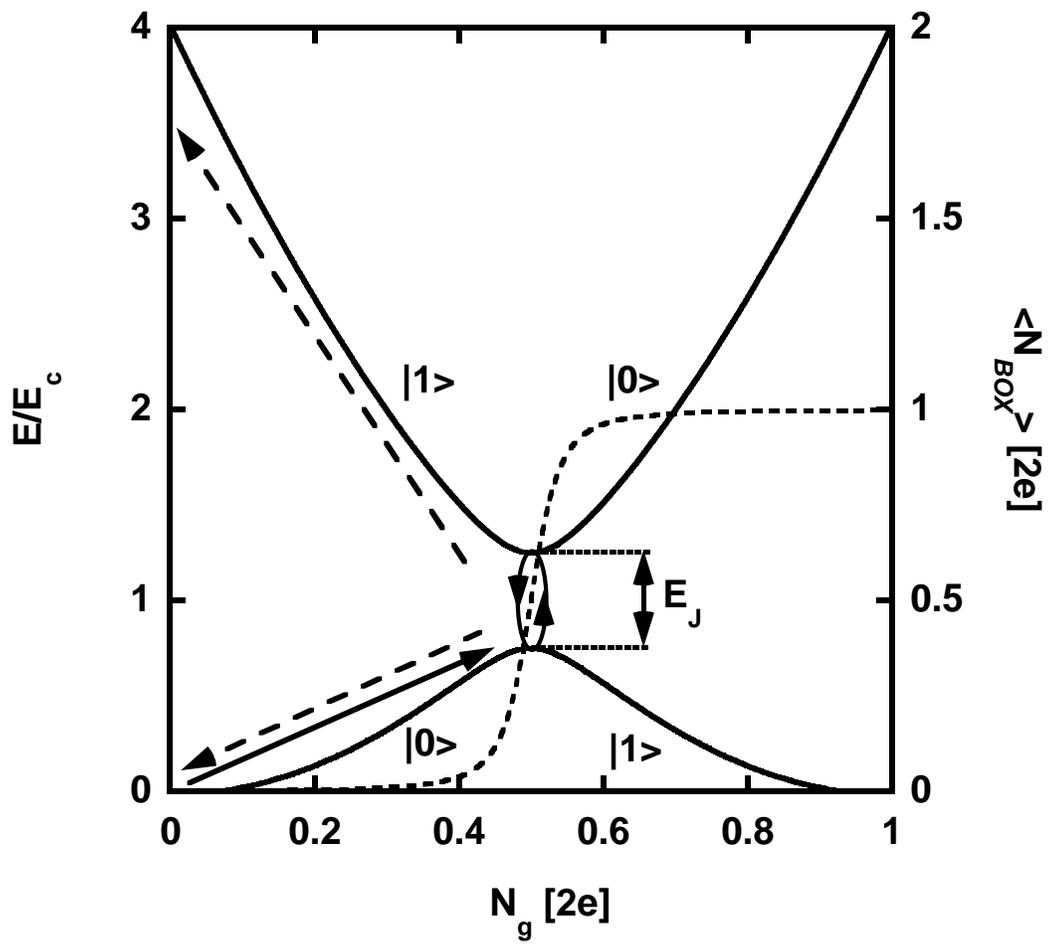

**Figure 2.**

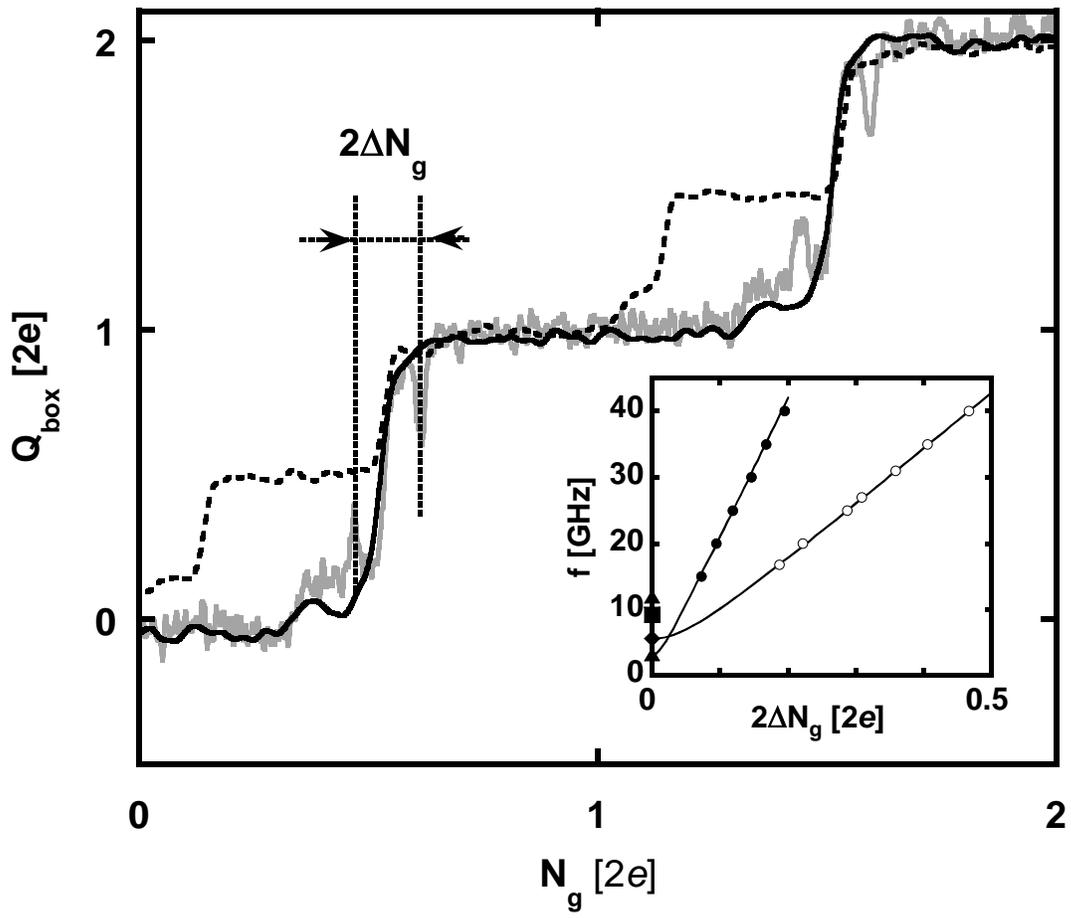

**Figure 3.**

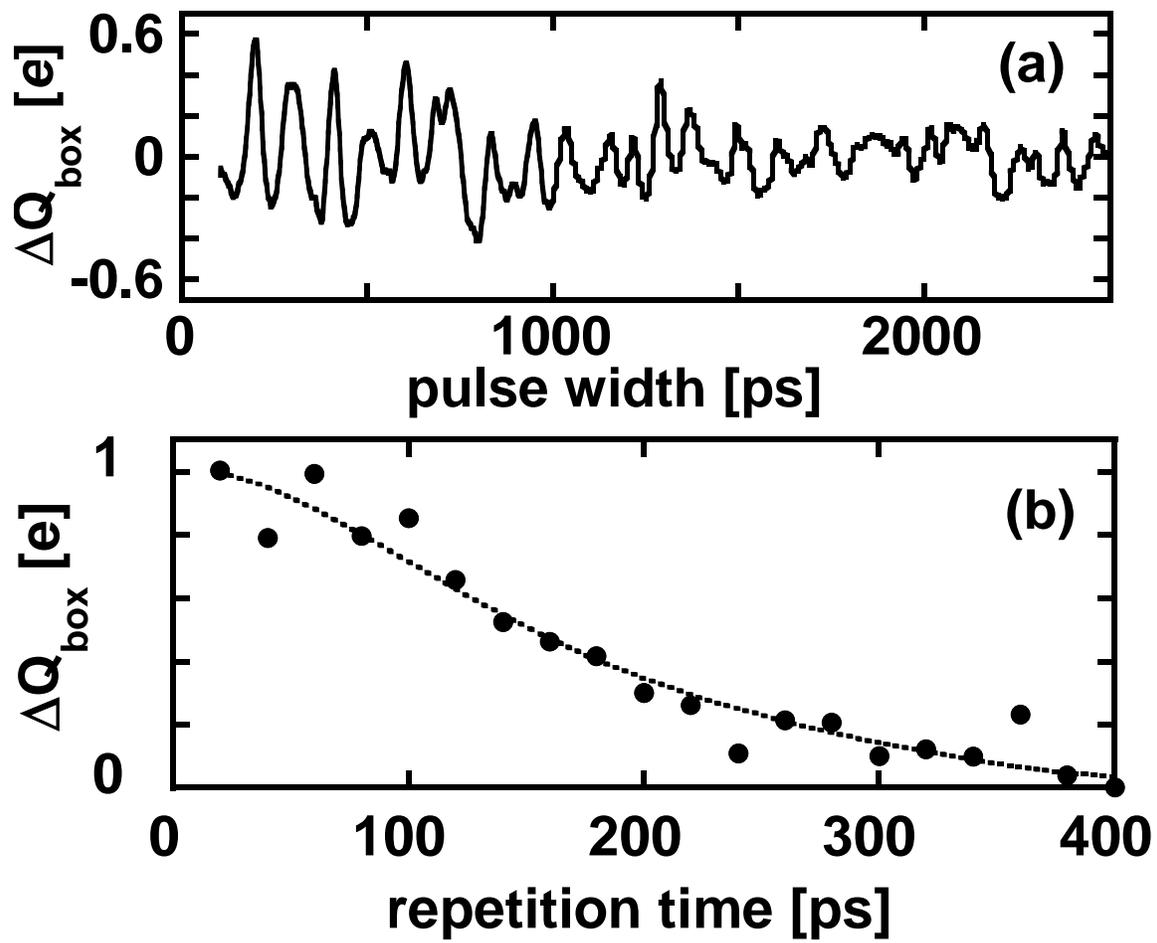

**Figure 4.**